\documentclass{aa}  

\usepackage{graphicx}
\usepackage{txfonts}
\usepackage{lipsum}
\usepackage{subcaption}         
\usepackage{lscape}             
\usepackage{placeins}           
                                
\usepackage[colorlinks=true,allcolors=blue]{hyperref}
\usepackage{fancyhdr}
\usepackage{soul}
\usepackage{balance}
\usepackage{enumitem}
\usepackage[dvipsnames]{xcolor}

\usepackage{xspace}
\newcommand{\kms}{km~s$^{-1}$\xspace}

\newcommand{\Msun}{M$_{\odot}$\xspace}
\newcommand{\agebar}{age$_{\rm lw,\,bar}$\xspace}
\newcommand{\agedisk}{age$_{\rm lw,\,disc}$\xspace}

\newcommand{\metbar}{log(Z/Z$_{\odot}$)$_{\rm lw,\,bar}$\xspace}
\newcommand{\metdisk}{log(Z/Z$_{\odot}$)$_{\rm lw,\,disc}$\xspace}

\newcommand{\gal}{EGS-24154\xspace}
\newcommand{\oiii}{[\ion{O}{III}]$\lambda$5008\xspace}
\newcommand{\nii}{[\ion{N}{II}]$\lambda$6585\xspace}
\newcommand{\niid}{[\ion{N}{II}]$\lambda \lambda$6550,6585\xspace}

\newcommand{\ha}{H$\alpha$\xspace}
\newcommand{\hb}{H$\beta$\xspace}
\newcommand{\barolo}{\textsc{$^{\rm 3D}$Barolo}\xspace}
\newcommand{\degree}{\textsc{$^{\circ}$}\xspace}
\newcommand{\vout}{\textsc{$v_{\rm out}$}\xspace}
\newcommand{\vn}{\textsc{$v_{\rm n}$}\xspace}
\newcommand{\vb}{\textsc{$v_{\rm b}$}\xspace}

\begin{document}

   \title{The onset of stellar bars at Cosmic Noon} 
   \subtitle{Bar-driven quenching and AGN co-evolution in a mature disc galaxy}

    \titlerunning{Bar-driven quenching and AGN co-evolution in a mature disc galaxy at Cosmic Noon}

%

   \author{L. Costantin\inst{1}\fnmsep\thanks{Corresponding author: lcostantin@cab.inta-csic.es}
   \and L. Morelli\inst{2}\fnmsep\thanks{Corresponding author: lorenzo.morelli@uda.cl}
   \and C. Cabello\inst{1}
   \and V. Cuomo\inst{3}
   \and M. Perna\inst{1}
   \and J.~A.~L. Aguerri\inst{4,5}
   \and M. Roshan\inst{6}
   \and E. Borsato\inst{3,7}
   \and E.~M. Corsini\inst{7,8}
   \and A. de Lorenzo-Cáceres\inst{4,5}
   \and F. D'Eugenio\inst{9,10}
   \and D. Gasparri\inst{2}
   \and A. Habibi\inst{6}
   \and T. Kim\inst{11,12}
   \and Y.~H. Lee\inst{13,14,15}
   \and J. Méndez-Abreu\inst{4,5}
   \and A. Pizzella\inst{7,8}
   \and B. Rodríguez Del Pino\inst{1}
    }

   \institute{Centro de Astrobiolog\'ia (CAB), CSIC-INTA, Ctra. de Ajalvir km 4, Torrej\'on de Ardoz, E-28850, Madrid, Spain
   \and Instituto de Astronom\'a y Ciencia Planetaria, Universidad de Atacama, Avenida Copayapu 485, Copiap\'o, Chile
   \and Departamento de Astronom\'ia, Universidad de La Serena, Av. Ra\'ul Bitr\'an 1305, La Serena, Chile
   \and Instituto de Astrof\'isica de Canarias, Calle V\'ia L\'actea s/n, 38205 La Laguna, Tenerife, Spain
   \and Departamento de Astrof\'isica de la Universidad de La Laguna, Avda.~Astrofis\'ico Francisco S\'anchez S/N, E-38206 La Laguna, Tenerife, Spain
   \and Department of Physics, Faculty of Science, Ferdowsi University of Mashhad, PO Box 1436, Mashhad, Iran
   \and INAF – Osservatorio Astronomico di Padova, Vicolo dell’Osservatorio 5, 35122 Padova, Italy
   \and Dipartimento di Fisica e Astronomia ``G. Galilei'', Universit\`a di Padova, vicolo dell’Osservatorio 3, 35122 Padova, Italy
   \and Kavli Institute for Cosmology, University of Cambridge, Madingley Road, Cambridge CB3 0HA, UK
   \and Cavendish Laboratory -- Astrophysics Group, University of Cambridge, 19 JJ Thomson Avenue, Cambridge CB3 0HE, UK
   \and Department of Astronomy, Yonsei University, Seoul 03722, Republic of Korea   
   \and Department of Astronomy and Atmospheric Sciences, Kyungpook National University, Daegu 41566, Republic of Korea
   \and Astronomy Program, Department of Physics and Astronomy, Seoul National University, 1 Gwanak-ro, Gwanak-gu, Seoul 08826, Republic of Korea
   \and SNU Astronomy Research Center, Seoul National University, 1 Gwanak-ro, Gwanak-gu, Seoul 08826, Republic of Korea
   \and Department of Astronomy and Atmospheric Sciences, Kyungpook National University, Daegu 41566, Republic of Korea
   }

    \date{Received \today}

  \abstract
  {Observations with the JWST revealed an unexpected abundance of barred 
  galaxies at Cosmic Noon. However, the physical properties of these early 
  bars are almost unconstrained, as it is their impact in the structural
  evolution of high-redshift disc galaxies.}
   {We derived the stellar populations of \gal, a barred spiral galaxy at $z=1.17$. 
   First, we investigated the role of the stellar bar in the early 
   assembly history and structural evolution of the galaxy.
   Second, we studied the properties of the interstellar medium to shed 
   light on the interplay with the central supermassive black hole.}
   {We analysed medium-resolution ($R\sim1000$) NIRSpec/IFS data of \gal 
   through full-spectral fitting and derived light and mass-weighted ages 
   and metallicities. We then reconstructed the spatially-resolved 
   star formation history, derived the ionizing mechanisms of the 
   interstellar medium analysing \hb, \oiii, \ha, and \nii emission lines, 
   characterized the ionized gas dynamics of \gal, and constrained the properties of a biconical outflow launched by the active galactic nucleus (AGN).}
   {\gal is a baryon-dominated, gas-rich disc galaxy, which grew more 
   than 90 per cent of its stellar mass when the Universe was $\sim2$~Gyr old. 
   We found that the stellar population of the bar started to form at $z\sim5$, 
   compatibly to the time when the stellar disc started to assemble. 
   We observed a \emph{star formation desert} in the bar region, 
   which is responsible for quenching star formation over several Gyr. 
   We then interpreted that the feedback from the AGN 
   prevented the growth of central mass concentration, 
   allowing the stellar bar to grow in size and strength.}
   {In this first study of spatially-resolved stellar populations 
   of a barred disc galaxy at $z>1$, we demonstrated how stellar bars 
   are key drivers of the early structural and dynamical evolution of disc galaxies. 
   In particular, our results call for a revision of most models 
   of disc and bar formation in early baryon-dominated, gas-rich disc galaxies.}


   \authorrunning{L.~Costantin et al.}

   \maketitle

\section{Introduction}

Observations with the James Webb Space Telescope 
\citep[JWST;][]{Gardner.J:2023}
opened a new window to study the structure of galaxies at $z>1$ 
\citep{Jacobs.C:2023, Boogaard.L:2024, Pandya.V:2024}, 
revealing an abundance of disc galaxies even up to $z>3-5$
\citep{Kartaltepe.J:2023, HuertasCompany.M:2025, Costantin.L:2025}.
Interestingly, an increasing number of studies revealed 
that early discs host stellar bars and spiral arms
\citep{Costantin.L:2023, Guo.Y:2023, Guo.Y:2025, LeConte.Z:2024, 
Geron.T:2025, Amvrosiadis.A:2025, Huang.S:2025, Wang.X:2026},
with a decreasing fraction from 
$\sim60$ per cent at $z=0$ \citep{Eskridge.P:2000, 
Barazza.F:2008, Aguerri.A:2009, Nair.P:2010, Masters.K:2011},
to $\sim18$ per cent at $1 < z < 2$ 
and $\sim14$ per cent at $2 < z < 3$ \citep{LeConte.Z:2024}.
These remarkable findings not only doubled the 
number of barred galaxies at $z>1$ compared to
previous studies \citep[e.g.,][]{MenendezDelmestre.K:2007, 
Sheth.K:2008, Melvin.T:2014, Simmons.B:2014},
but also push the redshift frontier of morphological diversity
to more than 11 Gyr ago, 
when bar-driven internal evolution started.

This recent landscape paved the route for revisiting
classical theories about the formation and 
early evolution of disc galaxies. In particular,
\citet{vanderWel.A:2025} proposed that galaxies experienced
an early ($z>2-3$), accelerated ($\sim$ few Myr), 
internal secular evolution ($\sim$10-30 orbital periods),
explaining the rapid build-up of a broad variety 
of morphologies at Cosmic Noon ($z=1-3$).
This framework complements the findings of
\citet{BlandHawthorn.J:2023}, which suggested that 
barred galaxies formed efficiently and
fast at high redshift in a baryon-dominated potential
\citep[see also][]{BlandHawthorn.J:2024, Fragkoudi.F:2025}.
Contrary to classical expectations 
\citep{Sellwood.J:1993, Bournaud.F:2005, Kraljic.K:2012, Reddish.J:2022},
these high-z bars could be long-lived in gas-rich, turbulent environments 
\citep{BlandHawthorn.J:2024, BlandHawthorn.J:2025, Huang.S:2025}, 
as recently confirmed at the highest
redshift by the detection of a stellar bar in GN20,
a disc galaxy at $z=4.05$ with an extreme gas fraction
\citep[$\sim 70\%$;][]{Boogaard.L:2026}.

Despite the recent effort of detecting barred galaxies
beyond $z\sim0$, most of the information is limited
to photometric surveys. Thus, spectroscopic information is
key to shed light on the early phases of galaxies growth, their structural diversity, and the impact that
different dynamical components -- including stellar bars --
have in shaping the Universe as we see it today.

From local spectroscopic studies, we know that stellar bars 
are fundamental dynamical drivers of late, secular evolution of disc galaxies
\citep{SanchezBlazquez.P:2014, Gadotti.D:2019, FraserMcKelvie.A:2020}.
They are responsible for the internal morphological evolution, 
facilitating the exchange of 
angular momentum from the inner
to the outer regions and the dark matter halo
\citep{Sellwood.J:1981, MartinezValpuesta.I:2006}.
They efficiently funnel gas inwards,
promoting the formation of central mass concentrations
\citep{Pfenniger.D:1990, Cheung.E:2013, Pastras.S:2025, Pastras.S:2026, Jolly.JB:2026},
nuclear structures \citep[e.g., discs, lenses, rings;][]{Kormendy.J:1979, 
Buta.R:1995, Corsini.E:2003, Coelho.P:2011, 
deLorenzoCaceres.A:2019, Gadotti.D:2020, LeConte.Z:2026},
and feed the growth of active galactic nuclei 
\citep[AGN;][]{Shlosman.I:1990, Combes.F:2019,  
SilvaLima.L:2022, Garland.I:2024, Combes.F:2026}.

Stellar bars are an efficient channel for the
global quenching of star formation 
in late stages of galaxy evolution
\citep{Masters.K:2012, Haywood.M:2016, Khoperskov.S:2018, Scaloni.L:2024}.
Indeed, stellar bars could play a fundamental role
in the transformation of galaxies
between the blue cloud and the red sequence
\citep{Strateva.I:2001, Baldry.I:2004}
and an even more important role in driving the
fast quenching of the high-redshift population 
of quiescent galaxies
recently identified with JWST \citep[e.g.,][]{Carnall.A:2023, deGraaff.A:2025}.
In particular, \citet{James.P:2016} quantified that 
the almost complete suppression of star formation
in the inner region of nearby barred galaxies corresponds
to the radial range swept out by the bar, 
a region that they termed \emph{star formation desert}
\citep[see also, e.g.,][]{James.P:2018, Neumann.J:2020}.
The bar induces strong shocks and shearing gas motions
which prevent the gas from cooling and collapsing
to efficiently form stars \citep[][but see also \citealt{Verley.S:2007}]{Reynaud.D:1998}.

At Cosmic Noon, the conditions 
of the interstellar medium (ISM)
were quite different from $z\sim0$.
Galaxies were more gas-rich
\citep{Tacconi.L:2020, ForsterSchreiber.N:2020}
and quite turbulent 
\citep{Genzel.R:2006, Ubler.H:2019, Wisnioski.E:2019}.
However, no spectroscopic, spatially-resolved studies
linking the stellar and ISM build-up and interplay 
in disc galaxies are available at $z>1$ to date.
In this context, we present the first of such investigations,
deriving the stellar populations and ISM properties
of \gal, a barred galaxy at $z=1.17343 \pm 0.00071$.
which was one of the first barred galaxies
identified with JWST \citep[][]{Guo.Y:2023}.

This is the second paper of a series of works focusing
on the stellar properties of \gal, as part of the GO3 program 5766 
\citep[PIs: V.~Cuomo, M.~Roshan, J.A.L.~Aguerri;][]{Cuomo.V:2024jwst}.
In particular, in \citet[][hereafter Paper~I]{Aguerri.A:2026}, 
we present the photometric analysis of \gal and EGS-12823,
building on the early results of \citet{Guo.Y:2023},
and present their spatially-resolved kinematics
of the gas and stellar components,
the first of such kind of observations at $z>1$.
In \citet[][hereafter Paper~III]{Cuomo.V:2026}, we derived
the bar pattern speed of \gal and discuss the implications
of such measurements at Cosmic Noon.

In this work, we time the build-up of the bar and disc, measuring the
light and mass-weighted age and metallicity as well as the enhancement 
of [Mg/Fe] of both structural components.
We finally explore the role of the bar in quenching
the central star formation and the impact of the supermassive
black hole in the evolution of the galaxy.
The paper is organized as follows. In Sect.~\ref{sec:section2}, 
we describe the data. In Sects.~\ref{sec:section3} and \ref{sec:section4},
we derive the stellar population properties of the galaxy 
and those of the ISM,
respectively. In Sects.~\ref{sec:section5} and 
~\ref{sec:section6}, we discuss our findings and
provide our conclusions.

Throughout this work, we assumed vacuum emission
line wavelengths and a cosmology 
with $H_0 = 67.4$~km~s$^{-1}$~Mpc$^{-1}$, 
$\Omega_{m} = 0.315$, and $\Omega_{\Lambda} = 0.685$
\citep{Planck:2020}. 
At the redshift of \gal, 
1~arcsec corresponds to 8.5~kpc.
All errors are reported as the 16th–84th percentile interval,
unless stated otherwise.


\section{Data} \label{sec:section2}

\gal (RA=214.8743~deg, DEC=+52.8869~deg; J2000)
was observed with NIRSpec integral field spectroscopy (IFS) on March 1, 2025
\citep[GO3-5766;][]{Cuomo.V:2024jwst}. 
The IFS observations were taken with 
the grating/filter pair G140M/F070LP. 
This results in a data cube with nominal spectral resolution 
$R\sim1000$ over the observed wavelength range $0.90-1.27$~$\mu$m
(but see below).
In Fig.~\ref{fig:figure1}, we show a composite image
of \gal combining NIRCam/F115W (optical rest-frame), 
NIRCam/F200W (near-infrared rest-frame),
and \ha monochromatic image from NIRSpec.

As described in Paper~I,
the raw data were calibrated with a modified version of the
JWST Science Calibration pipeline version 1.15.0 \citep{Bushouse.H:2023}
under the CRDS context \texttt{jwst\_1299.pmap}. 
The custom steps are extensively described 
in \citet{Perna.M:2023} and \citet{D'Eugenio.F:2024}.
The observations were combined on a cube with a 
pixel scale of 0.09~arcsec using the \texttt{drizzle} algorithm. 
Since the error extension of the calibrated cube 
is known to underestimate the actual errors,
we followed the prescription detailed in
\citet{Ubler.H:2023} and rescaled it 
to match the noise measured in regions 
with no contamination from any source.

We detected that the calibrated cube
is contaminated by a strong 
emission line at $\lambda_{\rm rest}\sim4980$~\AA,
identified as a metastable He line
\citep[$\lambda_{\rm obs} = 1.083~\mu$m; 
see][]{Brammer.G:2014, Arribas.S:2024}.
This line allowed us to measure on the data the
instrumental resolution of the observations.
Indeed, we extracted spectra from each 
spaxel and model the He line with a Gaussian component.
As a result, we obtained an estimation
of the instrumental resolution at each spatial position
of the cube, which is used to correct the kinematics maps
(see Sect.~\ref{sec:section4} and Paper I).
The median spectral resolution is of $R=749^{+75}_{-66}$,
which corresponds to an instrumental 
resolution of $\sigma_{\rm inst} = 170 \pm 16$~\kms.
Since the metastable He line is present 
over the full detector, and lies between \hb and \oiii,
we removed its contribution from the calibrated cube.
This was achieved by deriving a smoothed median spectrum 
from a $5\times5$ px$^2$ background region on the 
West side of the galaxy
and correcting each individual spaxel of the cube.

\begin{figure}[t!]
\includegraphics[width=0.48\textwidth]{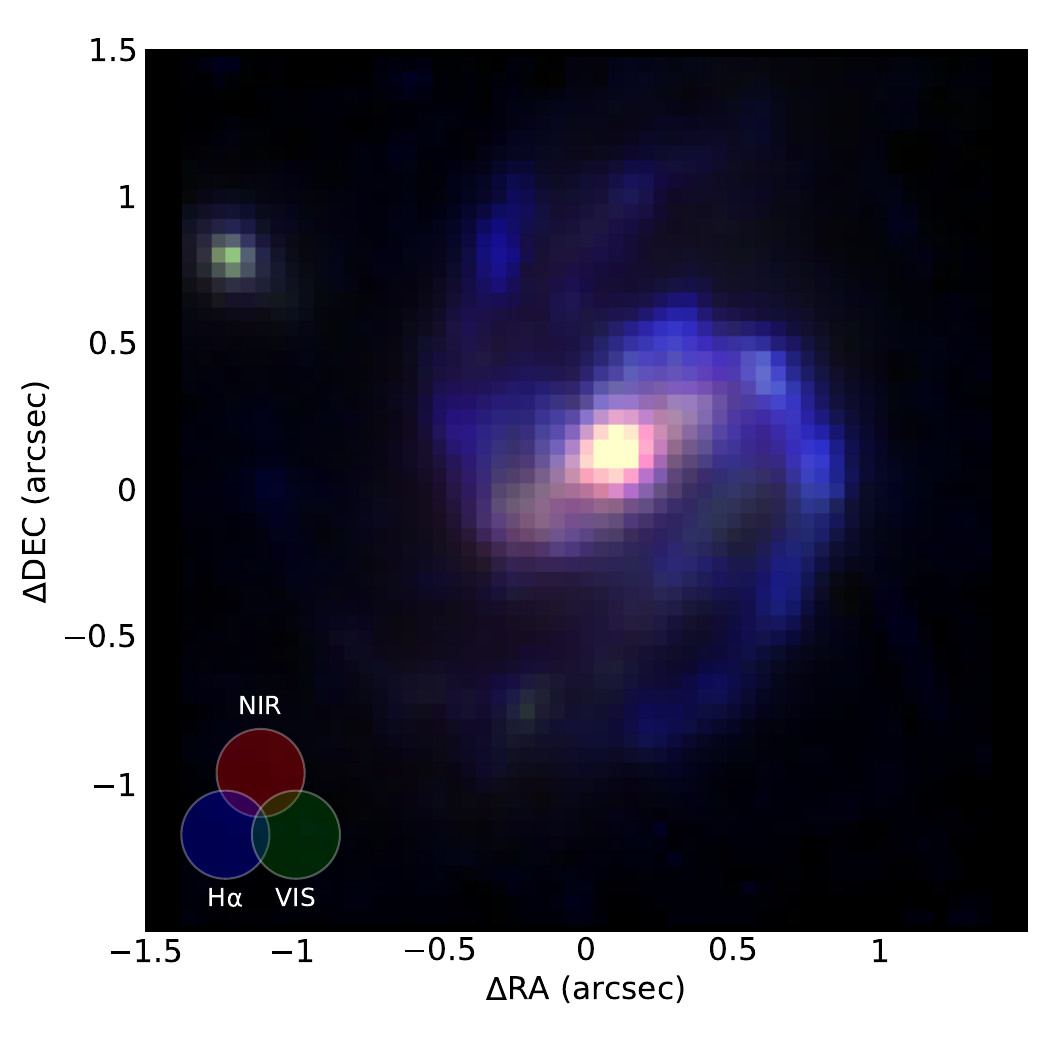}
\caption{Composite RGB image of \gal.
The colour image was built combining two NIRCam filters, F115W and F200W, 
and the \ha monochromatic image from the NIRSpec cube
(all resampled at the same pixel scale  of 0.05~arcsec~px$^{-1}$).
At $z=1.17$, F115W and F200W trace the rest-frame optical 
and near-infrared portion of the spectral energy distribution,
respectively. The field of view is oriented with North up and East left.
\label{fig:figure1}}
\end{figure}

A visual inspection of the raw data revealed that there
are multiple spectral features readily observable on the detector
but outside the nominal wavelength range,
including \ha and \niid lines.
Following the strategy employed in \citet{D'Eugenio.F:2025},
we therefore extracted the data by extrapolating 
the flat-field curves and wavelength solution beyond 
the nominal range, up to $1.45$~$\mu$m.
We finally checked that the error on the wavelength solution 
is negligible, comparing the kinematics of H$\alpha$ 
to other Balmer lines (see Sect.~\ref{sec:section3}).
It is worth noting that this version of the data cube
was used to calculate \nii/\ha
line ratios (Sect.~\ref{sec:section4}), so that 
an absolute flux calibration is not necessary.

We binned the final datacube employing a convex 
tessellation using the \textsc{PowerBin} algorithm 
\citep{Cappellari.M:2025} with target $S/N=25$ per bin,
ideal to reliably measure the stellar populations 
of \gal \citep[e.g.,][]{Gallazzi.A:2009}.
We measure the signal and the noise in the spectral 
window $\lambda_{\rm rest} = 4400-4700$~\AA, where 
the stellar component is less contaminated by nebular emission.
We excluded all spaxels with $S/N < 7$, to further improve
the quality of our spectra.
After the tessellation, the cube consists of
66~unique regions ($329$ individual spaxels),
covering $\sim2.5\times1.5$~arcsec$^2$,
corresponding to $\sim21\times13$~kpc$^2$.
Finally, 11 bins in the outer region 
were discarded because the final $S/N$ was not
sufficient to retrieve a reliable fit.


\section{Stellar populations \label{sec:section3}}

\subsection{Spatially-resolved ages and metallicities \label{sec:section3.1}}

\begin{figure*}[ht!]
\includegraphics[width=\textwidth]{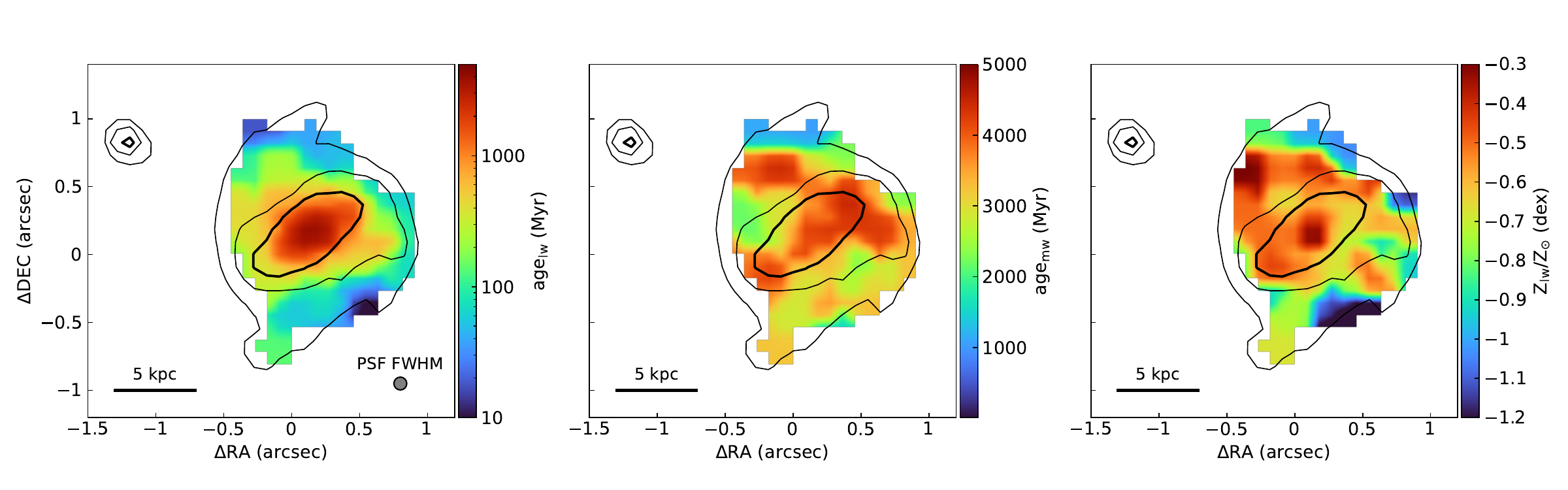}
\caption{Stellar population properties of \gal.
Light-weighted stellar age (left panel), 
mass-weighted stellar age (middle panel), and
light-weighted stellar metallicity (right panel).
The innermost contour defines the isophote
of the bar region (thicker black line; 
semi-major axis length of 3.6~kpc, 
as identified in \citet{Guo.Y:2023}). 
NIRSpec IFS PSF FWHM of $\sim$0.1~arcsec 
\citep[][]{D'Eugenio.F:2024} is reported.
Maps are shown with bilinear interpolation over
the tessellation.
\label{fig:figure_stpop}}
\end{figure*}

We analysed each spectrum with 
the Penalized Pixel-Fitting algorithm 
\citep[\textsc{pPXF};][]{Cappellari.M:2017, Cappellari.M:2023},
modelling both the stellar and gas components simultaneously.
The stellar component was fitted as a (non-negative) 
linear superposition of simple stellar-population (SSP) spectra 
matching the spectral resolution of the NIRSpec observations.
For the stellar templates, 
we employed the synthetic library of SSP from 
Flexible Stellar Population Synthesis 
\citep[FSPS;][]{Conroy.C:2009, Conroy.C:2010}.
The library uses \citet{Salpeter.E:1955} initial mass function 
and MESA Isochrones and Stellar Tracks \citep{Choi.J:2016}. 
We considered models in a 2D age-metallicity logarithmic grid,
spanning 38 age bins from 1~Myr to 5~Gyr and eight [M/H] bins 
from $-1.75$ to $0$~dex. In particular, the age grid is limited
by the age of the Universe at $z=1.173$, plus a buffer of 200~Myr.

To retrieve the mass-weighted age and metallicity of \gal,
we followed a procedure similar to that 
employed in \citet{Ikhsanova.A:2025}:
\begin{enumerate}
\item We normalized each spectrum by the median flux 
per spectral pixel to penalize non-smooth weight distributions
and achieve regularized solutions 
\citep[see][for all details]{Cappellari.M:2017}.
\item We fitted each spectrum including both 
additive (\texttt{degree=5}) and multiplicative 
(\texttt{mdegree=9}) polynomials to retrieve the best
parametrization of the galaxy stellar kinematics 
(i.e., velocity and velocity dispersion). 
In this first iteration, no regularization was included
and outliers were removed using a 10$\sigma$ clipping.
\item At this stage, we fixed the kinematics and 
iteratively fitted each spectrum with no polynomials 
but different regularization factors,
until we achieved a desired $\chi^2$ $\sim$ degrees of freedom.
\end{enumerate}
The full-spectral fitting was performed in the 
wavelength range $\lambda_{\rm rest} = 4180 - 5800$~\AA,
relying on several spectral features 
(i.e., H$\gamma$, \hb, Mgb, G-band, Fe4383, Fe5270, Fe5335)
to derive the stellar populations of \gal.
We derived both light (SDSS $g$ band) and
mass-weighted ages and metallicities.
Errors are estimated measuring galaxy properties 
from 100 perturbations of each spectrum with the noise.

In Fig.~\ref{fig:figure_stpop} (left and middle panels), 
we show the spatially-resolved
light-weighted and mass-weighted age of \gal.
On one side, ($g$-band) light-weighted ages 
are dominated by the brightest, younger stars 
and thus trace the most recent episodes of star formation; 
on the other side, mass-weighted ages 
average over the entire stellar population and better 
represent the mass assembly history of the galaxy
\citep{,serra:2007,zibetti:2017,Costantin.L:2019}. 

Focusing on light-weighted ages,
we found that the bar region has older stellar populations
than the surrounding disc, with median 
\agebar $= 1.6^{+1.5}_{-0.6}$~Gyr and 
\agedisk $= 106^{+248}_{-52}$~Myr, respectively.
We extracted the radial profiles of light-weighted ages
along multiple position angles, finding an overall 
negative gradient. Ages peaks at 
age$_{\rm lw} = 3.8$~Gyr in the central region
(see e.g., Fig.~\ref{fig:figure_stpop}, left panel)
and are at least $\sim1$~Gyr old up to $r \sim 3$~kpc
along the major axis and 
up to $r \sim 2$~kpc along all other directions.
The main age drop along the major axis corresponds
to the position of the bar ansae at the bar-disc interface,
also consistent with prominent of \ha and \hb emission
in those regions.

Focusing on mass weighted ages,
we found that \gal appears older on average
with respect to light-weighted estimations, having 
median age$_{\rm mw} = 3.2^{+1.0}_{-0.9}$~Gyr 
and age$_{\rm lw} = 467^{+171}_{-414}$~Myr.
In the bar region, we found 
median age$_{\rm mw,\,bar} = 3.9^{+0.3}_{-0.7}$~Gyr, 
which would correspond to a formation redshift $z\sim4.2$.
The stellar populations of the disc have similar ages
(median age$_{\rm mw,\,disc} = 3.1^{+0.9}_{-0.6}$~Gyr,
but some regions formed earlier at $z\sim6$),
which suggests that the galaxy formed the bulk of its
stars on short timescales,
but also dynamically settled
its morphology in multiple structural components early on
(see Sect.~\ref{sec:section3.2}).

\begin{figure*}[ht!]
\includegraphics[trim=2cm 0cm 2cm 0cm, width=\textwidth]{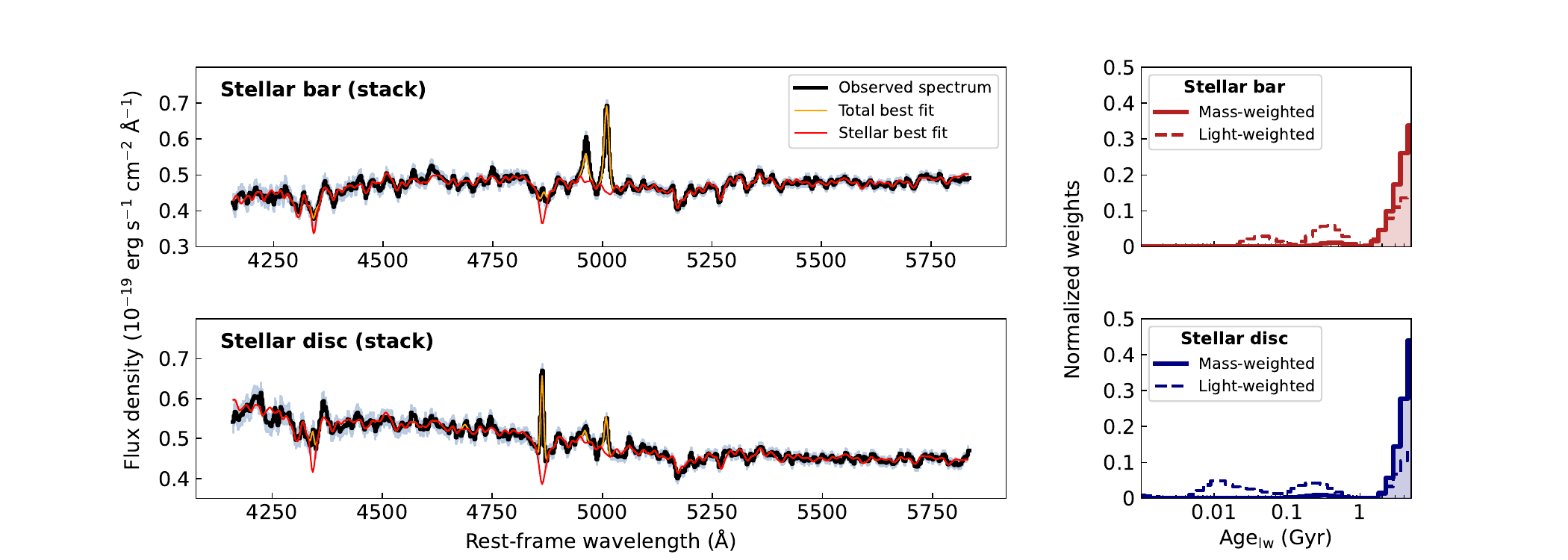}
\caption{Stack spectra and best fit models
of different morphological regions of \gal (left panels)
and the corresponding age
distributions, marginalized over metallicity (right panels). 
Mass and light-weighted stellar age distributions are
shown as filled and empty histograms, respectively.
\label{fig:figure3}}
\end{figure*}

The structural assembly history of \gal is confirmed 
by the light-weighted metallicity map 
(Fig.~\ref{fig:figure_stpop}, right panel).
At face values, the bar region is more metal-rich than
the surrounding disc, but still compatible within errors,
with \metbar = $-0.52^{+0.06}_{-0.14}$~dex 
and \metdisk = $-0.7^{+0.2}_{-0.3}$~dex, respectively.
Moreover, estimating the Mg/Fe-enhancement
from Lick spectral indices 
\citep{Worthey.G:1994, Worthey.G:1997, Trager.S:1998},
we derived a median value of $0.38^{+0.12}_{-0.08}$~dex
(consistent in the inner and outer region).
This result suggest an efficient early 
assembly (and fast quenching of
star formation) in $\sim100-300$~Myr 
\citep{Thomas.D:2005, Thomas.D:2010, Renzini2006}.
This result, if taken at face value, 
could be at odd with local bars being less Mg/Fe-enhanced
that their surrounding discs \citep[see e.g.,][but 
see \citealt{deLorenzoCaceres.A:2019}]{Neumann.J:2020}.
This raises a note of caution
in comparing galaxies across cosmic time. 
Secular, slow processes (e.g., prolonged or continuous 
formation of stars and/or orbital mixing) 
could be really effective in timescales of several Gyr
(as from $z=1$ to $z=0$), but could bias the interpretation
of the actual timescales for building-up 
these structures at the highest redshift.

\subsection{Stack spectra of different structural components
\label{sec:section3.2}}

We complemented the spatially-resolved stellar 
population analysis by creating stack spectra 
of different structural regions of \gal,
i.e., the stellar disc and stellar bar
(Fig.~\ref{fig:figure3}).
For the latter, we considered 
all spaxels within the inner isophote 
shown in Fig.~\ref{fig:figure_stpop}.
We corrected for the different 
line-of-sight velocities in different spatial regions
and stacked individual
bins using inverse-variance weighting.
We fitted the bar and disc 
spectra with \textsc{ppxf},
as described in Sect.~\ref{sec:section3.1},
and derived light-weighted ages (and metallicities)
for each individual component.
We found that star formation, as traced by \hb,
is prominent in the stellar disc and almost absent in
the bar region. 
The stellar disc shows three clear
stellar populations: a very young one
($\sim$10-30~Myr), an intermediate one 
($\sim$100-700~Myr), and an old one
($\sim$2-4~Gyr),
all of them with similar contributions in light.
The stellar bar still presents 
three stellar populations, but milder
contribution from the younger ones.
In particular, the light is dominated by a mature 
population of age $\sim$1-4~Gyr.
Combining these results with the spatially-resolved
analysis presented in Sect.~\ref{sec:section3.1},
we found that the intermediate-age population
of the bar starts to contribute more
moving towards the main 
star-forming knots at the end of the bar,
at the interface with the ansae.
Thus, we interpreted these results
as indicating that the stellar bar is 
an efficient driver of star formation quenching. 
As a consequence, we put a lower limit on the time 
of the bar build-up, assuming that most of 
the pre-existing disc stars where trapped 
into bar-supporting orbits
at the latest $\sim$2~Gyr back in time, 
corresponding to $z\sim2$.

\section{Co-evolution with the supermassive black hole and baryonic dominance in \gal \label{sec:section4}}

\subsection{Ionizing structure of \gal \label{sec:section4.1}}

We derived the fluxes of multiple emission lines
to characterize the spatial distribution of the ISM.
The \hb flux was measured with \textsc{ppxf},
thus using the values corrected for stellar absorptions
(see Sect.~\ref{sec:section3}).
As for \ha and \niid, we fitted the observed
(non flux-calibrated) spectrum with a model of three 
single Gaussian functions \citep[\textsc{lmfit};][]{Newville.M:2025}
for each emission line in all bins of the datacube.
For the \niid doublet we tied the ratio of the flux
to the theoretical value of 1:3 and we assumed that \niid and
\ha have the same velocity dispersion.
No assumptions were made for the other emission lines
and \ha is not corrected for the stellar absorption.
The \oiii line presents a second, blueshifted kinematic component
in multiple spaxels in the central region of the galaxy
(see an example in Appendix~\ref{sec:appendixB}).
Thus, we modelled these spaxels with a two-component
Gaussian fit to account for the blueshifted 
wings in \oiii, while the remaining spaxels
were modelled with a single Gaussian component. 
We noticed that the second component was not present
in \hb, while the spectral resolution of the observations
does not allow us to draw conclusions about the \ha profile,
which is partially blended with \niid lines.

\begin{figure*}[ht!]
\centering
\includegraphics[width=0.95\textwidth]{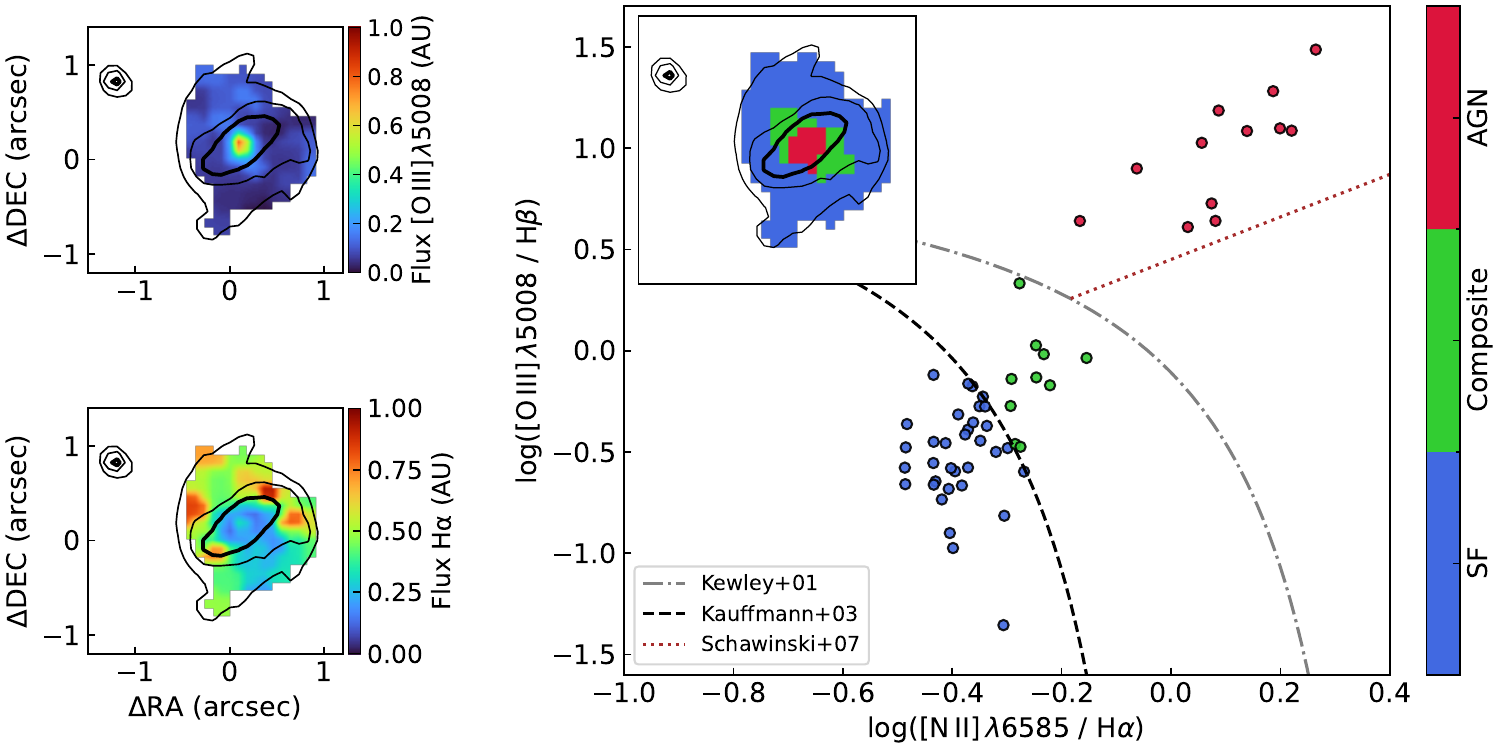}
\caption{Maps of \ha and \oiii normalized fluxes 
(top and bottom left panels, respectively).
BPT diagram of individual tessellation regions
of \gal (right panel). 
The gray dash-dotted line marks the theoretical 
maximum starburst line derived by \citet{Kewley.L:2001}. 
The black dashed line separates starburst galaxies and AGNs
\citep{Kauffmann.G:2003}. 
The red dotted line shows the demarcation between
Seyfert and low-ionization nuclear emission-line region 
galaxies \citep{Schawinski.K:2007}.
The inset panel shows the corresponding classification
overlaid on NIRCam F200W contours.
\label{fig:figure4}}
\end{figure*}

In Fig.~\ref{fig:figure4} (left panels), we mapped 
the normalized \oiii and \ha flux distribution.
On one side, \oiii peaks in the central region of \gal,
mainly confined in the inner $\sim$1.5~kpc. On the other side, 
\ha is more extended, with two bright knots
in correspondence of the bar ansae
and the NW spiral arm.
The same trend is present on the \hb distribution.
Interestingly, in the bar region, \ha is highly suppressed,
but it is enhanced in the inner $\sim$0.8~kpc.
From the \ha map, we cannot detect any filaments
connecting \gal to the companion galaxy 
(NW in Fig.~\ref{fig:figure1}), 
suggesting that there is no recent interaction in place,
at least in the latest few Myr.
Indeed, at the time of observations, we estimated
that the galaxies are probably gravitationally bound, 
but \gal is dynamically stable against the tidal influence
of the companion, that is, there is not 
dynamical perturbation ongoing
(see also Sect.~\ref{sec:section5} and Appendix~\ref{sec:appendixA}).

In Fig.~\ref{fig:figure4} (right panel), 
we showed the spatially-resolved 
BPT diagram \citep{Baldwin.J:1981} of \gal.
The diagram allowed us to distinguish 
the dominant contribution to ionization 
across the entire galaxy.
In particular, we see that star
formation dominates gas ionization in the
disc region and in the spiral arms, 
while the AGN dominates the inner galaxy region.
We found composite ionization
on both sides of the bar, which could be interpreted as 
due to AGN ionization cones.
Interestingly, the region of composite ionization 
is perpendicular to the major axis of the bar
(which is close to the major axis of the disc;
see Paper I).
We also found that \ha 
velocity dispersion increases along the same direction, 
with a PA of $\sim30^{\circ}$.
These results could be consistent 
with ionized gas preferentially escaping along 
lower-density directions
(see also Sect.~\ref{sec:section4.2} and
Appendix~\ref{sec:appendixB}), 
as usually observed in galactic outflows 
\citep[][]{Veilleux.S:2005}.

To complement the analysis detailed above, 
and shed further light on the ISM conditions 
and AGN properties of \gal,
we checked for complementary multi-wavelength observations
and found no X-ray detection \citep{Nandra.K:2015}.
However, \citet{Tacconi.L:2013} reported 
CO 3–2 detection from ancillary observations 
with the IRAM Plateau de Bure millimeter interferometer.
From these observations, they estimated
the cold molecular gas mass of 
\gal to be 
$M_{\rm mol} = 4.6 \times 10^{10}$~M$_{\odot}$
(systematic uncertainty of 50\%),
under the assumptions of $R_{13} = 2$ and 
$\alpha_{\rm CO} = 4.36$.
Then, assuming a stellar mass 
$M_{\star} = (1.29^{+0.99}_{-0.56}) \times 10^{11}$~M$_{\odot}$,
as measured in \citet[][]{Stefanon.M:2017}, 
we derived a gas fraction 
$f_{\rm gas} = M_{\rm mol} / (M_{\rm mol} + M_{\star}) = 0.25 \pm 0.13$, 
consistent with typical 
main-sequence galaxies at $z\sim1$ \citep{Tacconi.L:2020}.

\subsection{\ha disc model and dark matter content
\label{sec:section4.3}}

To estimate the mass budget of \gal,
we modelled the \ha best-fit IFS cube 
with \barolo \citep{DiTeodoro.E:2015}
to derive the circular rotational velocity.
While \barolo assumes that the galaxy has a 
geometrically thin disc and the kinematics is
dominated by pure rotational motion, 
radial motions can also be modelled.
With an additional radial component
we aim at reproducing the 
S-shaped twist in the iso-velocity contours
\citep{vanderKruit.P:1978, Pizzella.A:2004, Price.S:2021},
being aware that this is clearly a simplified assumption
to account for the presence of the bar
\citep[see also][]{Ubler.H:2024}.

\begin{figure*}[ht!]
\includegraphics[trim=0cm 0.5cm 0cm 0.5cm, width=\textwidth]{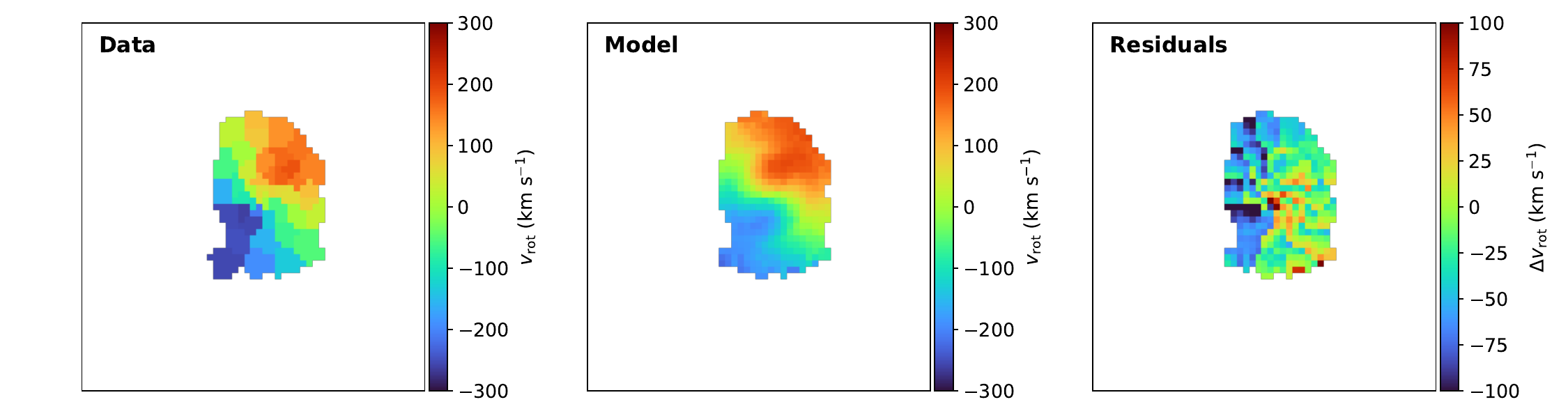}
\caption{Best-fit \barolo model of the \ha velocity field in \gal,
including both rotational and radial motions.
From left to right: observed, modelled, and residual
moment-1 maps.
\label{fig:figure5}}
\end{figure*}

As in \citet{Perna.M:2026}, we followed a 
three steps strategy
\citep[see also][for all details]{DiTeodoro.E:2021, Perna.M:2022}.
In summary, we run multiple
azimuthal models with different inclinations
$i_{\rm phot}=37^{\circ} \pm 7^{\circ}$ 
(using steps of $1^{\circ}$), 
where $i_{\rm phot}$ is the inclination derived from the 
isophotal analysis of the galaxy surface brightness, 
assuming an infinitesimally thin disc
(see Paper~I).
In this first runs, we fixed the inclination
while we minimized over three free parameters: the rotational velocity $v_{\rm rot}$, the velocity dispersion $\sigma$, and the major axis PA.
Comparing the $\chi^2$ of each run, 
we found the best-fit inclination $i_{\rm kin}=38^{\circ}$.
Thus, we run \barolo with a local flux normalization
to minimize $v_{\rm rot}$, $\sigma$, $i_{\rm kin}$,
and the major axis PA.
This second run provides a first-order model of \gal,
but fails in the inner region,
possibly because of the bar-like (and spiral) velocity field.
Therefore, we run \barolo with a radial component 
in addition to the rotational motions.
The best model for the velocity field
and the residuals for this latter run
are shown in Fig.~\ref{fig:figure5},
and display a clear S-shaped pattern 
as the one detected in data.

Using \ha as a tracer, 
we found that the best-fit model of \gal
has $i_{\rm kin} = 38\degree$ and mean major axis PA
on the receding half of the galaxy
of $337^{\circ}$.
At a radial distance of 8~kpc, 
we derived a rotational velocity,
corrected for inclination, $v_{\rm rot} = 312$~\kms.

We estimated the dynamical mass of \gal as
\begin{equation}
M_{\rm dyn} = \frac{v_{\rm circ}^2 \, r}{G},
\end{equation}
where $r$ is the outer radius where we 
resolved both the stellar and gas kinematics ($r=8$~kpc), 
$G$ is the gravitational constant, and
$v_{\rm circ}^2$ is the circular velocity.
In particular, by assuming the ionized gas 
to be pressure supported 
\citep[e.g.,][]{Bertola.F:1995, Cinzano.P:1999},
we derived the circular velocity as
\begin{equation}
v_{\rm circ}^2(R) = v_{\rm rot}^2(R) - \sigma^2(R) \left[ \frac{d\ln\Sigma(R)}{d\ln R} + \frac{d\ln\sigma^2(R)}{d\ln R} \right],
\end{equation}
where $v_{\rm rot}(R)$ is the rotation velocity 
of the ionized gas derived from 
3D kinematic modelling (i.e., \barolo)
and $\sigma(R)$ is the intrinsic velocity 
dispersion of the H$\alpha$ emitting 
gas corrected for instrumental broadening.
The surface density term $\Sigma(R)$ is not directly 
measured due to the lack of flux calibration of 
the H$\alpha$ data. Instead, we adopt the H$\alpha$ 
surface brightness $F_{\rm H\alpha}(R)$ as a proxy, 
assuming $\Sigma(R) \propto F_{\rm H\alpha}(R)$. 
This proportionality cancels in logarithmic derivatives, 
allowing the use of $d\ln F_{\rm H\alpha}/d\ln R$ 
in place of $d\ln \Sigma/d\ln R$.
The second term accounts for the radial variation 
of the velocity dispersion field, 
traced directly from the observed 
$\sigma_{\rm H\alpha}(R)$ map.

Under these assumptions, at a radial distance of 8~kpc,
we derived $v_{\rm circ} = 336$~\kms and
$M_{\rm dyn} = 2.1 \times 10^{11}$~\Msun.
Thus, considering a baryonic mass of 
$M_{\rm baryons} = M_{\star} + M_{\rm mol} = 1.75^{+1.02}_{-0.61}\times10^{11}\ {\rm M_\odot}$
and a dynamical mass derived 
considering bar-like radial motions,
we found that, within the inner 8~kpc, 
\gal is baryon-dominated with a fraction
of dark matter of only $\sim 17\%$.

It is worth noticing that
the value of the circular velocity is
similar to the one derived
as $v_{\rm circ, 0}^2 = v_{\rm rot}^2 + k\sigma^2$ = 320~\kms,
where $k=2$ is an approximation
for pressure support (asymmetric drift) 
in dynamically warm, turbulent disc galaxies
\citep[][]{ForsterSchreiber.N:2009, Burkert.A:2010, Genzel.R:2017}.
\newline

\subsection{AGN feedback \label{sec:section4.2}}

\begin{figure*}[ht!]
\includegraphics[width=\textwidth]{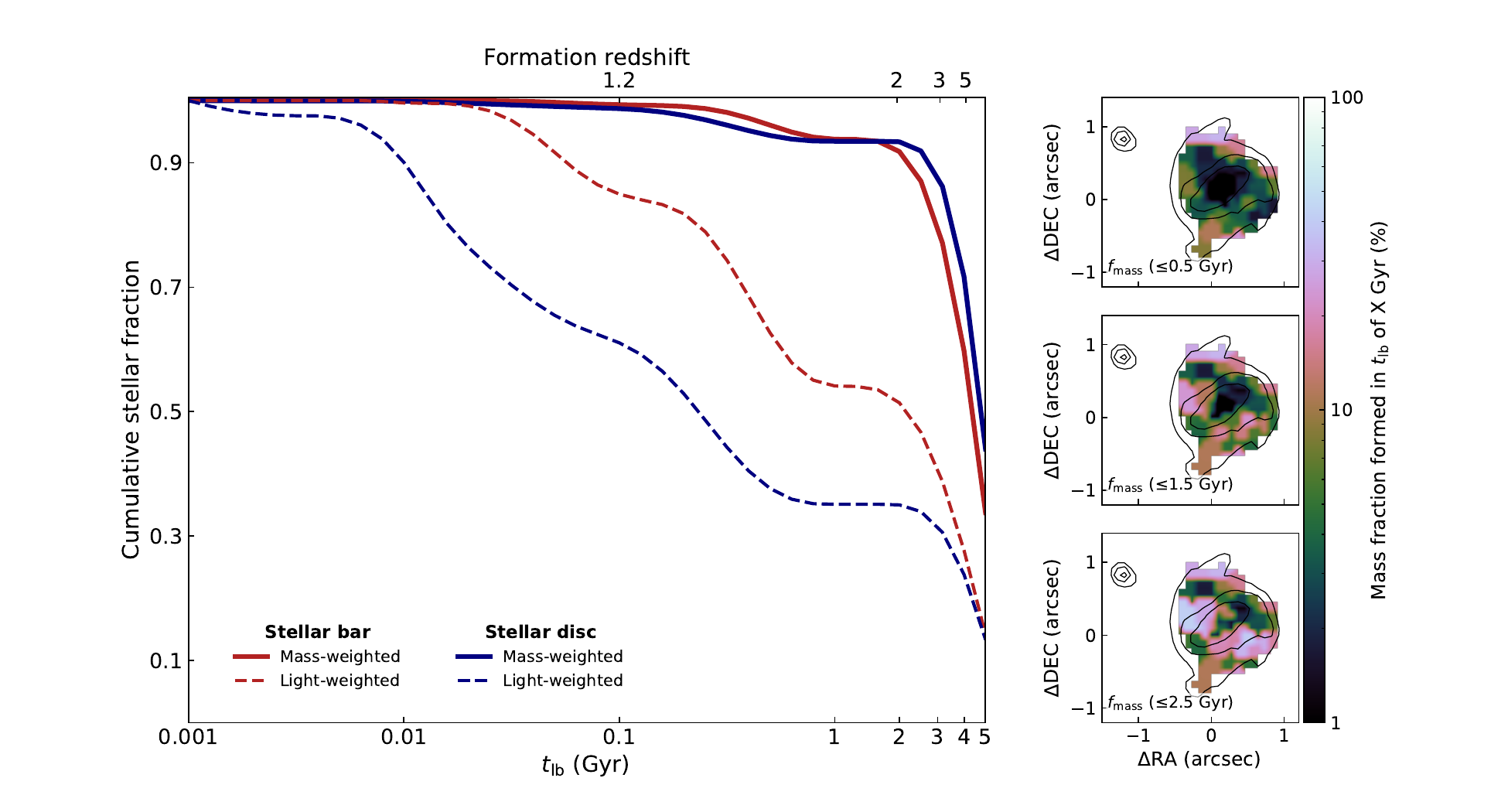}
\caption{Left panel: cumulative stellar fraction as
a function of $t_{\rm lb}$, the lookback time from $z=1.17$.
It represents the fraction of mass (solid lines)
and light (dashed lines) that is older than
a given age for the stellar bar and disc
(shown in red and blue, respectively).
Right panels: spatially-resolved maps of the fraction of mass 
formed in a lookback time $t_{\rm lb}$ = 1, 2.5, and 4~Gyr.
\label{fig:figure6}}
\end{figure*}

In this section, we characterized the feedback
effects of the AGN. 
In the following, given the low spatial resolution and the
unknown detailed geometrical configuration of the outflow, the energetics 
were derived assuming a simple single radius wind. 
Because of this, we considered for the outflow 
velocity a median value derived from 
the pixels were a second Gaussian component 
is required to reproduce the \oiii line profile.
Indeed, as described in Sect.~\ref{sec:section4.1} and Appendix~\ref{sec:appendixB},
\oiii presents two components with different kinematics:
a first component, tracing the
systemic velocity of the \gal, and a second component,
tracing the outflowing ionized gas. 
In particular, the outflow
is aligned perpendicular to the bar major axis,
and it extends up to $\sim 3-4$~kpc
(see Fig.~\ref{fig:figure6}).

Assuming that the second, blueshifted \oiii component 
is tracing the AGN-driven outflow, we derived 
the outflowing gas velocity as
\begin{equation}
\vout = |\vb - \vn| + \dfrac{{\rm FWHM_{\rm b}}}{2} \,,
\end{equation}
where |\vb - \vn| is the velocity shift 
between the broad (second) and narrow (first) components
and ${\rm FWHM_{\rm b}}$ is the full width at half maximum
of the broad component, corrected for the 
instrumental dispersion 
\citep[see][]{Xu.Y:2025, RodriguezdelPino.B:2026}.
Thus, we measured a median
outflowing gas velocity of $\vout = 891^{+81}_{-392}$~\kms.

Assuming an isothermal gravitational potential
\citep{Heckman.T:2000, Veilleux.S:2005, Xu.Y:2025},
we derived the escape velocity 
\begin{equation}
v_{\rm esc}(r) =
v_{\rm circ}\,
\sqrt{2\left[1+\ln\left(\frac{r_{\rm max}}{r_{\rm out}}\right)\right]} \, ,
\end{equation}
where $v_{\rm circ}$ is the circular 
velocity of the galaxy, 
$r_{\rm out} = 3$~kpc is the galactocentric radius 
of the outflowing gas, and $r_{\rm max} = 100$~kpc 
is the truncation radius of the isothermal halo.
Under these simplified assumptions,
we found \vout/$v_{\rm esc} = 0.88$.
This suggests that the outflowing gas is unlikely 
to escape the gravitational potential of the galaxy
and may instead remain gravitationally bound, 
eventually being recycled into the galaxy.

Integrating the spatially-resolved SFH
within the central 3 kpc, we derived an average
star formation rate of
$\mathrm{SFR}_{50}=0.6~M_\odot~\mathrm{yr^{-1}}$ 
and $\mathrm{SFR}_{100}=0.5~M_\odot~\mathrm{yr^{-1}}$ 
over the last 50 and 100 Myr, respectively. Then, 
following \citet{Carniani.S:2024}, we computed 
the ionized gas mass in the outflow as
\begin{equation}
M_{\rm ion, out} = 0.8 \times 10^8 
\left ( \dfrac{L_{[\ion{O}{III}]}}{10^{-44} \, {\rm erg~s^{-1}}} \right ) 
\left ( \dfrac{500 \, \rm cm^{-3}}{{n_e}} \right ) 
\left ( \dfrac{Z_{\odot}}{Z} \right ) 
\, {\rm M_{\odot}} \, ,
\end{equation}
where $L_{[\ion{O}{III}]}$ is the luminosity of the \oiii broad
component, $n_e$ is the outflow electron density,
and $Z$ is the outflow metallicity. 
In particular, we assumed that the ionized 
gas has solar metallicity
and that the outflow has
the same electron density as the one derived from 
the [\ion{S}{II}]$\lambda \lambda$6718,6733 
narrow component line ratio in the nuclear region of \gal
($n_e = 370$~cm$^{-3}$; see also 
\citealt{ForsterSchreiber.N:2019} and
\citealt{Parlanti.E:2025}).
Since we have no access to flux-calibrated \ha/\hb
flux ratio (and H$\gamma$ is not detected in the nuclear region; Fig.~\ref{fig:figure3}),
we cannot correct for extinction
the measured \oiii flux. Thus,
we derive a lower limit for the ionized gas mass
of $M_{\rm ion, out} > 1.9 \times 10^4 $~M$_{\odot}$.
Assuming that the mass outflow rate is constant with time,
we derived the mass outflow rate at a galactocentric radius
of 3~kpc of $\dot{M}_{\rm out, [\ion{O}{III}]\lambda5008} = 5.7$~M$_{\odot}$~yr$^{-1}$
\citep[e.g.,][]{Lutz.D:2020}.
The corresponding mass-loading factors are
$\eta=\dot{M}_{\rm out}/\mathrm{SFR} = 9.6$ and $10.8$
for SFR$_{50}$ and SFR$_{100}$, respectively.
This result suggests that the ionized outflow 
removes gas at a rate one order of magnitude 
higher than it is converted into stars.

Using the prescription of \citet{Netzer.H:2019},
we finally measured the AGN bolometric luminosity 
$L_{\rm bol}=5.2\times10^{43}~\mathrm{erg~s^{-1}}$
from the integrated H$\beta$ luminosity measured within $r=3$~kpc.
Then, following \citet{Perna.M:2019}, 
the corresponding kinetic power and momentum rate 
are $\dot{E}_{\rm kin}=1.4\times10^{42}~\mathrm{erg~s^{-1}}$, 
corresponding to
$\dot{E}_{\rm kin}/L_{\rm bol} \sim 3\%$,
and $\dot{P}_{\rm out}=3.2\times10^{34}~\mathrm{dyne}$, yielding a momentum boost of $\dot{P}_{\rm out}/(L_{\rm bol}/c) \sim 19$.

The derived energetics are consistent with those typically observed in AGN-driven ionized outflows \citep{Bertola.E:2025, Venturi.G:2025}.

\section{Discussion \label{sec:section5}}

\subsection{Stellar bar as main driver of SF quenching \label{sec:section5.1}}

The stellar population analysis revealed that 
\gal started to assemble as early as $z \sim 6$
(lookback time of $\sim$3-4~Gyr).
The stellar bar and stellar disc have consistent
mass-weighted ages, pointing to a rapid formation 
at early times.
Interestingly, star formation was heavily suppressed since
$z \sim 2-3$ in the region swept by the stellar bar.
This result confirms that the stellar bar 
assembled early on, since its stellar populations
have not changed, while the populations of the
stellar disc rejuvenated with time.
Indeed, light-weighted maps and \ha distribution shown 
in Figs.~\ref{fig:figure_stpop}-\ref{fig:figure4}
reveal that the stellar
disc has experienced more intense episodes of rejuvenation.
Among the possibilities for such rejuvenation,
we favour a scenario where pristine 
gas inflow from the halo sustained the SFR of the disc
\citep{Dekel.A:2009, SanchezAlmeida.J:2014}, 
but we cannot discard
a role of AGN feedback in driving gas  
from the nuclear region into the outer disc (Sect.~\ref{sec:section4.2}).
The increase in SFR could also
be triggered by the companion galaxy,
but it is unlikely given the poor tidal
influence that the latter had on \gal 
in timescales traced by \ha 
($<$10~Myr; see Appendix~\ref{sec:appendixA}).

As a complementary piece of evidence of the role of the 
stellar bar driving the quenching of star formation in \gal,
we derived the cumulative SFH by marginalizing 
the age weights obtained from \textsc{pPXF} over the
metallicity and integrating them across cosmic time
\citep[][]{GrebolTomas.P:2023, Ikhsanova.A:2025}. 
In Fig.~\ref{fig:figure6} (left panel), we separated 
the mass and light cumulative SFHs
of the stellar bar and disc,
highlighting the different evolution
of the stellar population of the two components.
In particular, we found that light-weighted stellar fractions
show the main differences in the history of the
two structural components.
The stellar disc is dominated by young stellar
populations dominating the light in the outer region,
despite representing only a small fraction of the stellar mass.
On the other hand, the stellar bar is systematically 
deficient in the youngest stellar populations.
From the comparison between light and 
mass-weighted stellar fractions,
we concluded that the stellar mass in the bar and disc 
formed during the same early epoch,
and the stellar bar has experienced 
little recent star formation.

To better quantify the spatial variation of the 
aforementioned trends,
we derived three different ``snapshots''
of the mass assembly history of \gal,
deriving the fraction of mass formed in a lookback time 
of 1, 2.5, and 4~Gyr (Fig.~\ref{fig:figure6}, right panels). 
We found that \gal formed most of its mass 
between 2.5 and 4~Gyr back in time ($z \sim 2.5-5$).
Indeed, the stellar disc started to assembly early,
since $\sim70\%$ of its mass was already formed
by $z=5$, as inferred from looking at
the mass accretion history in the outer region of \gal,
shown in Fig.~\ref{fig:figure6} (left and bottom right panels).
Stars from the disc where probably
trapped into bar-supporting orbits
at this stage, since only $<$10\% of stellar mass 
was formed in the last 2.5~Gyr in the bar region.
In the last Gyr,
almost no stars formed in the bar region (1-3\%),
a piece of supporting evidence of the role of the stellar bar
in driving the halt of star formation.

The role of the stellar bar driving the quenching
of \gal is also supported by the 
(ionized) gas distribution and kinematics.
As shown in Fig.~\ref{fig:figure4}, 
we detect abundant \ha and \hb along the spiral arms 
and at the bar-disc interface, a 
clear sign of ongoing star-formation.
However, \ha is scarce in the bar region,
but abundant again in the nuclear region
(central $\sim$1~kpc; along with \oiii emission), mainly ionized by the AGN.
Thus, we witness no total depletion of 
gas reservoir in the inner kiloparsecs,
which is typical of super massive black hole (SMBH) kinetic feedback 
\citep{DiMatteo.T:2005, ForsterSchreiber.N:2014, Cheung.E:2016}.
Interestingly, not depleting the gas reservoir points toward
a common path in the quenching mechanisms 
of \gal and the Milky Way \citep[see e.g.,][]{Haywood.M:2016}.
Complementary, as discussed in Sect.~\ref{sec:section4.3}
(see also Paper~I),
\gal shows high central velocity dispersion in the stellar and 
gas components ($\sigma_{0, \star} \sim 110$~km~s$^{-1}$
and $\sigma_{\rm 0, gas} \sim 180$~km~s$^{-1}$, respectively).
Shocks can effectively transform kinetic rotational energy 
into random gas motions, powering turbulence in the ISM.
As a consequence, the bar region becomes stable against gravitational
fragmentation, resulting in the halt of star formation
\citep[][]{Khoperskov.S:2018}.

As a result of the points discussed above,
we interpreted that star formation was
halted and further inhibited by the
influence of the stellar bar,
mainly along its major axis.
We also found a negative radial gradient
of light-weighted ages along the bar minor axis,
which could be interpreted as a 
\emph{star formation desert} in the region
swept by the stellar bar,
similarly to that found in observations 
of local barred galaxies
\citep{James.P:2016, James.P:2018, Neumann.J:2020},
and cosmological simulations
\citep[e.g., ErisBH,][]{Guedes.J:2011, Bonoli.S:2016, Spinoso.D:2017}.

\subsection{Stellar bar and AGN co-evolution \label{sec:section5.2}}

In Sect.~\ref{sec:section4},
we characterized the ISM properties of \gal
and found that the ionization of the nuclear region
is well-explained by the presence of an AGN.
However, the interplay between 
inflow gas streams driven by the stellar bar
and AGN outflow is quite complex.
In the context of cosmological simulations, 
in TNG50 \citep{Pillepich.A:2019, Nelson.D:2019},
\citet{Frosst.M:2025} found that 
92\% of barred galaxies at $z=0$ 
present 3-15 kpc wide gas holes in their centre
due to SMBH kinetic feedback, 
which ejects gas and suppress central star formation.
However, \gal is at odd with such behaviour,
suggesting a marginal role of AGN feedback in
suppressing central star formation,
at least at present.
Indeed, the outflow was not capable 
of depleting the molecular gas reservoir, 
as instead observed in many AGN systems at high redshift 
\citep{Perna.M:2018, Bertola.E:2025}.

While AGN feedback is (yet) not efficient 
in depleting the nuclear region
from gas, we consider it has a key role in
the evolution of \gal. Indeed,
AGN feedback has been shown to prevent
the grow of central mass concentration
\citep{Bonoli.S:2016, Kataria.S:2018}.
As a consequence, the stellar bar
can grow in size and strength,
as also shown in TNG50, when 
AGN feedback is enhanced \citep{Zana.T:2019}.
In our work, these results are confirmed by recent measurements
presented in Paper~I,
showing that \gal has a very long ($R_{\rm bar} = 5.9$~kpc)
and strong bar (amplitude of the $m=2$ 
Fourier mode $S_{\rm bar} = 0.45$).

This analysis shows the complex
interplay between stellar bar dynamics and
AGN feedback, unveiling the
efficient role of the first in
quenching star formation in \gal,
while the latter results critical
in sustaining bar growth across cosmic time.
High-resolution observations
of the cold molecular gas
could give us a more definitive picture of
the complex dynamical status of \gal,
providing fundamental constraints
for cosmological simulations of Milky Way-like
barred galaxies at Cosmic Noon 
\citep[e.g.,][]{Fragkoudi.F:2025, BlandHawthorn.J:2025}
and shedding light on the puzzling interplay 
between AGN feedback and bar-driven quenching 
in baryon-dominated, gas-rich barred galaxies.

\subsection{The onset of stellar bars at Cosmic Noon
\label{sec:section5.3}}

In this work, we have shown the \gal 
is a mature disc galaxy, 
with very old stellar populations
in the bar region, possibly forming
at $z\gtrsim3$ in short timescales.
The stellar bar is long-lived (several Gyr), 
despite the ionized gaseous disc is slightly turbulent
(see Paper~I, for more details),
with velocity dispersions of $\sigma_{\rm H\alpha} > 50$~\kms
in the disc region
and up to $\sigma_{0, \rm H\alpha} \sim 180$~\kms
in the central region.
These values are in agreement with
velocity dispersions measured in low-redshift 
galaxies of similar mass
\citep[e.g.,][but with important differences between
spirals and ellipticals]{Pizzella.A:2004, Ganda.K:2006, Yu.X:2019}
and also with the turbulence measured 
at Cosmic Noon for the ionized gas component
\citep[e.g.,][]{Johnson.H:2018}.
As a note of caveat, cold gas velocity dispersions could be
up to three times smaller compared to the ionized gas
component \citep[see e.g.,][]{Rizzo.F:2024},
calling for the need of both tracers to actually characterize
the dynamical status of disc galaxies.
In this context, the turbulence of the cold gas component
and the overall abundance of gas in disc galaxies 
at high redshift, is crucial for properly understanding the 
physical processes responsible for bar formation,
as pointed out in the recent theoretical works
of \citet{BlandHawthorn.J:2025} and observationally 
confirmed by \citet{Huang.S:2025} and \citet{Boogaard.L:2026}.
Indeed, these studies demonstrates that gas-rich 
discs support the fast formation of stellar bar 
in the early Universe, 
similarly to what we propose in this work,
motivating a new theoretical 
perspective on the formation of stellar bars 
in such extreme conditions.

Based on the result of Sect.~\ref{sec:section4.2}
and the arguments presented in \citet{Frosst.M:2026},
we tend to favour a scenario where the stellar bar
could have been triggered by a past tidal interaction
at $z > 2.5$ (up to $z\sim5-6$).
Indeed, \citet{Frosst.M:2026} reported
that baryon-dominated discs in TNG50 can secularly form
stellar bars under relatively weak tidal perturbation.
In Sect.~\ref{sec:section4.1}, we advocated 
how the tidal interaction between \gal and the 
companion galaxies was not responsible for triggering 
the recent rejuvenation of the stellar disc.
Thus, pockets of gas reservoir should have been present
in \gal, possibly in the nuclear region, as 
a consequence of bar-driven inflows.
Among other possibilities described in Sect.~\ref{sec:section4}, 
this concentration of gas
could have been moved outwards
from AGN-driven feedback only recently.

In a broader context, given the increasing 
number of barred galaxies observed (and expected)
at the highest redshift \citep[up to $z>3$;][]{Costantin.L:2023, Amvrosiadis.A:2025, Boogaard.L:2026, Perna.M:2026, Wang.X:2026} 
we speculate that stellar bars could have played a pivotal role
in early galaxy evolution. On one side, they could have been
the dynamical drivers of the fast growth of early bulges,
supposedly in place since $z\sim5-6$ \citep[e.g.,][]{Tacchella.S:2015, Costantin.L:2021, HuertasCompany.M:2024}.
On the other side, early stellar bars could have been
responsible for the fast mass growth and quenching 
of galaxies, as a consequence
of their efficiency in driving inflow of a massive amount
of turbulent gas and rearranging stars into the galaxy potential.
Thus, this structural transformation could
impact the observed downsizing of galaxies, 
with a fundamental role of gas dynamics in 
ruling the overall early assembly of morphological
diversity even beyond Cosmic Noon.

\section{Summary and conclusions} \label{sec:section6}

In this work, we derived with the stellar population properties
of \gal, a barred disc galaxy at $z=1.17$. 
We derived spatially-resolved ages and metallicities 
out to the stellar disc, finding that the 
galaxy started to form at $z \sim 5-6$, 
on timescales of 100-300~Myr.
We clearly distinguished different structural
components in the age maps. In particular, 
the stellar populations in the bar region 
formed on similar timescales as those of the 
stellar disc and has not experienced 
relevant star formation in a lookback time of, at least, 1~Gyr.
These results are confirmed by analysing the
stacked spectra of the bar and disc
regions.
Furthermore, bar-driven quenching results in a 
\emph{star formation desert}
in the region swept by the bar 
(i.e., negative age gradient along both the 
bar major an minor axis), remarkably similar 
to what observed in barred galaxies at low-redshift.

We also analysed the ionizing state of the interstellar
medium, finding that the nuclear region is dominated
by an AGN component. Furthermore, 
\gal shows composite ionization
in a direction perpendicular to the bar major axis,
which was then identified as an AGN-driven outflow
with broad, blueshifted \oiii emission.
Our analysis allows us to speculate that AGN feedback is 
essential to prevent the growth of a central mass concentration
and led the bar to grow in length and strength across cosmic time.
This results stresses the fundamental interplay
between bar-driven quenching and 
AGN-driven feedback in \gal.

We modelled the the \ha kinematics and derived that \gal
has a dynamical mass of $M_{\rm dyn} = 2.3 \times 10^{11}$~\Msun.
This allowed us to derive that \gal
is a baryon-dominated (70\% baryons over dark matter) disc galaxy,
with a gas fraction of $\sim25\%$.
This results has important implications
for the evolution of \gal, and especially for the
fast growth of the stellar bar. Indeed, (turbulent)
gas has been widely thought to prevent or, at least,
slow-down bar formation.
We witnesses the opposite trend, 
since we observed a long ($R_{\rm bar} = 5.9$~kpc) 
and strong ($S_{\rm bar} = 0.45$) bar, with 
stellar populations that formed 
in less than 300-500~Myr at $z \sim 4$.
Thus, gas dynamics was ruling the evolution
of \gal in the early epochs of its morphological
transformation, with (probably) a minor role
of stellar dynamics in shaping its structure
and mass assembly history.

All the previous results may need to be
complemented by high-resolution spatially-resolved
observations of the molecular gas component.
Indeed, we still miss key information on the cold gas
dynamics, its influence in the
growth of the supermassive black hole,
and in the overall evolution of \gal.

As shown by this series of work focusing on \gal 
(see also Paper~I, Paper~III),
JWST opened a new window in the understanding 
of the morphological diversity in the early Universe.
In particular, stellar bar are being proven
to be reliable clocks for unveiling the timescale
of galaxy growth as well as to explain
multiple physical phenomena observed at and beyond Cosmic Noon,
including the fast build-up and early quenching of galaxies.


\begin{acknowledgements}
We would like to thank H.~\"Ubler
for the support in the calibration of the 
data and valuable feedbacks.
This work is based on observations made with the
NASA/ESA/CSA James Webb Space Telescope. 
The data were obtained from the Mikulski Archive 
for Space Telescopes at the Space Telescope Science 
Institute, which is operated by the Association 
of Universities for Research in Astronomy, Inc., 
under NASA contract NAS 5-03127 for JWST. 
These observations are associated with program JWST-GO-5766.
The project that gave rise to these results 
received the support of a fellowship from 
the “la Caixa” Foundation (ID 100010434). 
The fellowship code is LCF/BQ/PR24/12050015. 
LC acknowledges support from grants 
PID2022-139567NB-I00 and PIB2021-127718NB-I00 
funded by the Spanish Ministry of Science 
and Innovation/State Agency of Research 
MCIN/AEI/10.13039/501100011033 and 
by “ERDF A way of making Europe”. 
VC acknowledges support by ANID 
through the FONDECYT grant nr.~11250723.
MP acknowledges support through the grants 
PID2021-127718NB-I00, PID2024-159902NA-I00, 
and RYC2023-044853-I, funded by the Spain Ministry 
of Science and Innovation/State Agency of 
Research MCIN/AEI/10.13039/501100011033 and 
El Fondo Social Europeo Plus FSE+.
JALA acknowledge support from the Agencia Estatal 
de Investigación del Ministerio de Ciencia, 
Innovación y Universidades (MCIU/AEI) under 
grant WEAVE: EXPLORING THE COSMIC ORIGINAL 
SYMPHONY, FROM STARS TO GALAXY CLUSTERS and 
the European Regional Development Fund (ERDF) 
with reference PID2023-153342NBI00/10.13039/501100011033.
AdLC acknowledges financial support from the 
Spanish Ministry of Science and Innovation 
(MICINN) through RYC2022-035838-I 
and PID2021-128131NB-I00 (CoBEARD project).
TK acknowledges support from the 
Basic Science Research Program through the 
National Research Foundation of Korea (NRF) 
funded by the Ministry of Education (No.~RS-2025-25399934).
YHL acknowledges support from the National Research
Foundation of Korea (NRF) funded by the Korean government 
(MSIT; No.~RS-2026-25482692) and by 
the Ministry of Education (No.~RS-2023-00249435).
BRDP acknowledges support from grants PID2021-127718NB-I00 
and PID2024-158856NA-I00 funded by Spanish Ministerio 
de Ciencia e Innovaci\'on MCIN/AEI/10.13039/501100011033 
and by “ERDF A way of making Europe”.
\end{acknowledgements}

\bibliographystyle{aa}

\begin{appendix}

\section{Companion galaxy \label{sec:appendixA}}

We estimated the effects of tidal interaction
between \gal and the companion galaxy 
located NW (see Fig.~\ref{fig:figure1}).

First, we measured the projected angular separation 
between the two galaxies, $\theta = 1.5$~arcsec, 
corresponding to a physical (projected) separation 
on the sky of $d_{\rm sky} \sim 13$~kpc. 
This represents the minimum possible distance between 
the galaxies.
To quantify the gravitational effect of the companion, we
computed the tidal parameter
\citep[see e.g.,][]{Henriksen.M:1996, Cortese.L:2007},
which measures the ratio of tidal
to internal gravitational acceleration:
\begin{equation}
T = 2 \frac{M_2}{M_{\rm dyn}} \left(\frac{r_1}{d_{\rm sky}}\right)^3 = 0.01 \, ,
\end{equation}
where $M_2 \sim 3.2 \times 10^{9}~M_\odot$ 
is the mass of the companion galaxy
and $r_1 \sim 10$~kpc is the radial extension of \gal.
This analysis indicates that tidal forces 
are at the percent level compared to 
the galaxy’s self-gravity and \gal 
is not affected by tidal stripping.

Second, we estimated the separation required 
for the companion to induce significant tidal perturbations. 
Inverting the tidal parameter, we found that 
the companion should have passed within $\sim4$~kpc to
significantly perturb the inner disc ($T>0.5$).
Assuming such a close passage occurred, 
we estimated the time since pericenter. 
For a perpendicular orbit and a disc inclination 
of $i = 37^\circ$, the deprojected distance is
$d_{3D} \sim 21$~kpc. Considering that the 
two galaxies have a relative velocity along the line of sight
$v_{\rm rel} \sim -256$~\kms 
($z_1 = 1.17343 \pm 0.00071$ 
and $z_2 = 1.17157 \pm 0.00001$, respectively),
the time since closest approach is 
$T_{\rm imp} \sim 100$~Myr \citep[see e.g.,][]{Perna.M:2026}.
This provides an order-of-magnitude estimate 
for the time since a hypothetical interaction 
capable of triggering star formation in \gal.

Finally, we considered the likelihood of such a close passage. 
The required pericenter ($\sim 4$ kpc) corresponds 
to a small geometric cross-section $\sigma \sim \pi R_{\rm peri}^2$.
Reaching such a small separation requires low angular momentum 
and a favourable orbital configuration. 
Given the low mass ratio ($>$ 1:50), 
the efficiency of tidal perturbations is further reduced.

In conclusion, despite the small projected separation, 
the current tidal influence of the companion is extremely weak
($T \sim 0.01$). 
A past close passage capable of inducing star formation 
cannot be ruled out, but would require a finely tuned 
orbit with pericenter $\lesssim 4$~kpc 
and a timescale of $\sim 100$ Myr. 
Such a scenario is possible but unlikely,
also considering that \ha traces star formation 
on $<10$~Myr timescales.

\section{Multi-component modelling of \oiii \label{sec:appendixB}}

In this Section, we provided additional details 
about the multi-component Gaussian model of
the \oiii emission line in the central region of \gal.

\begin{figure}[t!]
\includegraphics[trim=0cm 0.5cm 2.5cm 1.6cm, width=0.45\textwidth]{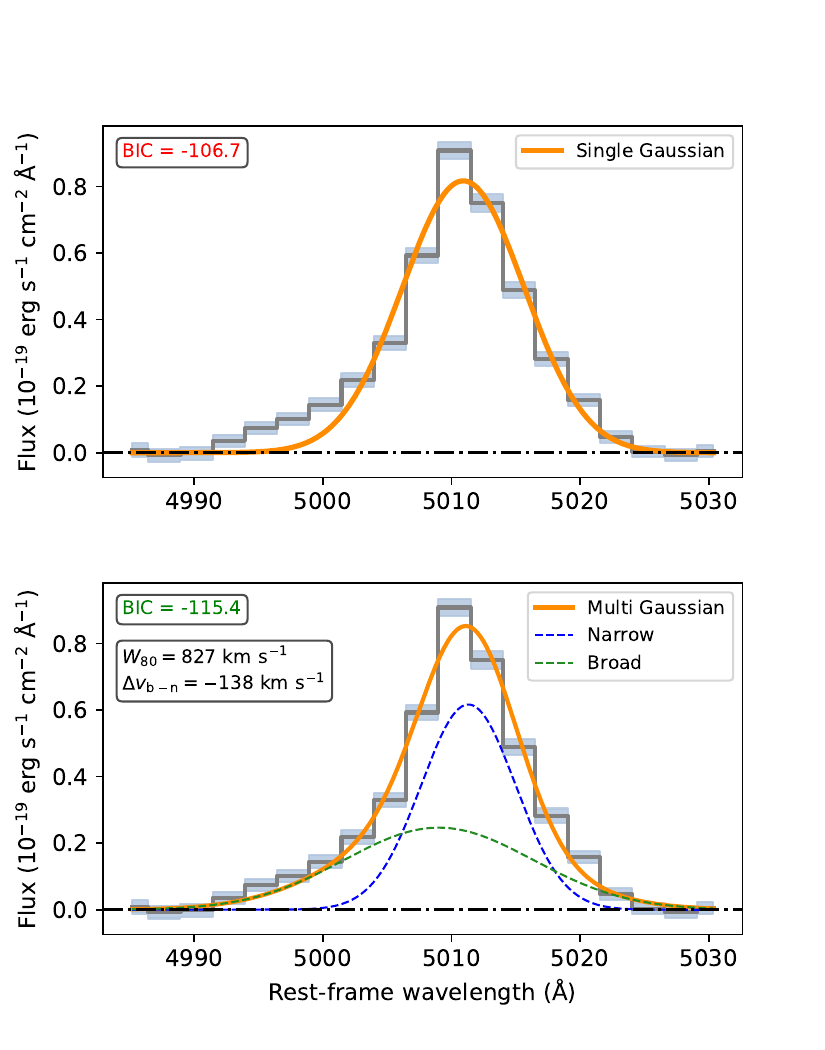}
\caption{Single (top panel) and multi-component model 
(bottom panel) of the \oiii emission line. 
The total model is shown as an orange solid line,
while the (first) narrow and (second) broad components are shown
as blue and green dashed lines, respectively.
We reported the value of the BIC statistics in each panel.
We also reported the difference in velocities 
between the broad and narrow components ($\Delta v_{\rm b-n}$)
and the width of the multi-component model ($W_{80}$).
\label{fig:figureB1}}
\end{figure}

In Fig.~\ref{fig:figureB1}, we reported an
example of the one-component (top panel) 
and two-component (bottom panel)
Gaussian fit of the emission line in a central spaxel
(marked in Fig.~\ref{fig:figure6}).
For the two-component Gaussian model, 
we identified the first component 
as the narrow \oiii line, tracing the systemic velocity of 
the ionized gas, and the second component as
a broader, blueshifted kinematic component 
tracing the outflowing gas.
As in \citet{Perna.M:2022}, we chose the number 
of kinematic components 
to model the \oiii line
by means of the Bayesian information criterion 
\citep[BIC;][]{Schwarz.G:1978}.

\begin{figure}[t!]
\includegraphics[trim=0.5cm 2.3cm 0.7cm 2.3cm, width=0.45\textwidth]{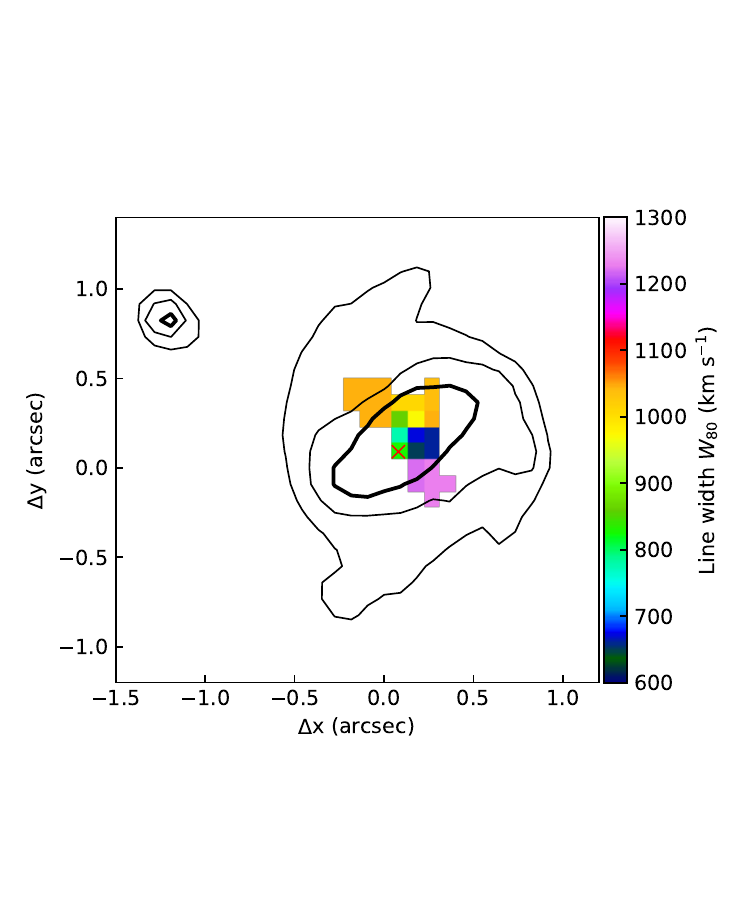}
\caption{Line width ($W_{80}$) map obtained 
from the two-component Gaussian fit.
A red cross marks the spaxel 
from which we extracted the spectrum 
shown in Fig.~\ref{fig:figureB1}.
\label{fig:figureB2}}
\end{figure}

Based on the multi-component line profile,
we measured the non-parametric velocities $v_{10}$ and $v_{90}$,
corresponding to the velocities at which 
10\% and 90\% of the flux is accumulated \citep[e.g.,][]{Liu.G:2013}. 
In particular, in Fig.~\ref{fig:figureB2} we showed 
the line width $W_{80} = v_{90} - v_{10}$ 
for all spaxels modelled with two kinematic components.
As mentioned in Sect.~\ref{sec:section4.1},
we found that the outflowing ionized gas 
is escaping preferentially along lower-density 
regions aligned with the minor axis of the bar,
out to a radius of $\sim 3-4$~kpc.
We also found that the \oiii line is broader in the
centre with respect to the outer part,
while being more blueshifted outside of the bar region.
Combined, these trends resulted in higher line widths
($W_{80} \sim 1100-1200$~\kms) in the outer regions 
of the outflow with respect to 
the centre ($W_{80} \sim 700-800$~\kms).

\end{appendix}

\end{document}